\documentclass[12pt]{article}
\usepackage[centertags]{amsmath}
\usepackage{txfonts}
\usepackage{epsfig}

\newcommand{\RR}{{\mathbb R}}
\newcommand{\CC}{{\mathbb C}}
\newcommand{\beq}{\begin{equation}}
\newcommand{\eeq}{\end{equation}}

\renewcommand{\Re}{\mathop{\mathrm{Re}}}

\newtheorem{lemma}{Lemma}[section]

\numberwithin{equation}{section}
\allowdisplaybreaks

\begin{document}
\title{An integral geometry lemma and its applications: the nonlocality of the
Pavlov equation and a tomographic problem with opaque parabolic 
objects}
\author{P.G.~Grinevich\thanks{L.D. Landau Institute for Theoretical Physics,
 Chernogolovka, Russia; Lomonosov Moscow State University, Moscow, Russia;
Moscow Institute of Physics and Technology, Moscow Region, Russia; 
E-mail: pgg@landau.ac.ru}, P.M.~Santini\thanks{
Dipartimento di Fisica, Universit\`a di Roma ``La Sapienza'', Roma, Italy; 
Istituto Nazionale di Fisica Nucleare, Sezione di Roma, Roma, Italy, 
E-mail: paolo.santini@roma1.infn.
it}}

\maketitle

\begin{abstract}
As in the case of soliton PDEs in 2+1 dimensions, the evolutionary form of 
integrable dispersionless multidimensional PDEs is non-local, and the proper 
choice of integration constants should be the one dictated by the associated 
Inverse Scattering Transform (IST). Using the recently made rigorous IST for 
vector fields associated with the so-called Pavlov equation 
$v_{xt}+v_{yy}+v_xv_{xy}-v_yv_{xx}=0$, we have recently esatablished that, in the 
nonlocal part of its evolutionary form 
$v_{t}= v_{x}v_{y}-\partial^{-1}_{x}\,\partial_{y}\,[v_{y}+v^2_{x}]$, the formal 
integral $\partial^{-1}_{x}$ corresponding to the solutions of the Cauchy 
problem constructed by such an IST is the asymmetric integral 
$-\int_x^{\infty}dx'$. In this paper we show that this results could be guessed 
in a simple way using a, to the best of our knowledge, novel integral geometry
lemma. Such a lemma establishes that it is possible to express the integral of
a fairly general and smooth function $f(X,Y)$ over a parabola of the $(X,Y)$ 
plane in terms of the integrals of $f(X,Y)$ over all straight lines non 
intersecting the parabola. A similar result, in which the parabola is replaced
by the circle, is already known in the literature and finds applications in 
tomography. Indeed, in a two-dimensional linear tomographic problem with a 
convex opaque obstacle, only the integrals along the straight lines 
non-intersecting the obstacle are known, and in the class of potentials  
$f(X,Y)$ with polynomial decay we do not have unique solvability of the 
inverse problem anymore. Therefore, for the problem with an obstacle, it is 
natural not to try to reconstruct the complete potential, but only some 
integral characteristics like the integral over the boundary of the obstacle. 
Due to the above two lemmas, this can be done, at the moment, for opaque 
bodies having as boundary a parabola and a circle (or, more generally, an 
ellipse). We expect that this result can be extended to a larger class of 
convex opaque bodies.          
\end{abstract}

\maketitle
\section{Introduction}

Integrable dispersionless PDEs in multidimensions, intensively studied in the 
recent literature (see \cite{MS6} for an account of the vast literature on 
this subject), arise as the condition of commutation $[L,M]=0$ of pairs of 
one-parameter families of vector fields. A novel Inverse Scattering Transform
(IST) for vector fields has been constructed, at a formal level in  
\cite{MS0}, \cite{MS1}, \cite{MS2}, \cite{MS3}, to solve their Cauchy problem,
obtain the long-time asymptotics, and establish if, due to the lack of 
dispersion, the nonlinearity is strong enough to cause a gradient catastrophe 
at finite time \cite{MS4}, \cite{MS5}. Due to the novel features of such IST 
(the corresponding operators are unbounded, the kernel space is a ring, the 
inverse problem is intrinsically non-linear), together with the lack of 
explicit regular localized solutions, it was important to make this IST 
rigorous, and this goal was recently achieved on the illustrative example of 
the so-called Pavlov equation \cite{Pavlov}, \cite{Ferapontov}, \cite{Duna} 
\beq 
\label{Pavlov}
\begin{array}{l}
v_{xt}+v_{yy}+v_xv_{xy}-v_yv_{xx}=0,~~v=v(x,y,t)\in\RR,~~x,y,t\in\RR, 
\end{array}
\eeq
arising in the study of integrable hydrodynamic chains  \cite{Pavlov}, and in 
Differential Geometry as a particular example of Einstein - Weyl metric 
\cite{Duna}. It was first derived in \cite{Duna1} as a conformal symmetry of 
the second heavenly equation. 

In the form (\ref{Pavlov}) it is not an evolution equation. To rewrite it in 
the evolution form, we have to integrate it with respect to $x$:
 \beq 
\label{Pavlov-ev}
v_{t}= v_{x}v_{y}-\partial^{-1}_{x}\,\partial_{y}\,[v_{y}+v^2_{x}],
~~v=v(x,y,t)\in\RR,~~x,y,t\in\RR, 
\eeq
where $\partial^{-1}_{x}$ is the formal inverse of $\partial_{x}$. Of course, 
it is defined up to an arbitrary integration constant, depending on $y$ and
$t$. On the other hand, the IST for integrable dispersionless PDEs provides 
us with a unique solution of the Cauchy problem
in which the function $v(x,y,0)$ is assigned, corresponding to a 
specific choice of such integration constant. 

In a recent paper we have specified the choice of the integration constant. 
More precisely, we have shown that the IST formalism for the Pavlov equation, 
a nonlinear analogue of the direct and inverse Radon Transform \cite{MS3}, 
corresponds to the following evolutionary form of the Pavlov equation:
\beq 
\label{Pavlov-ev2}
v_{t}(x,y,t)= v_{x}(x,y,t)\,v_{y}(x,y,t) +\int_{x}^{+\infty} 
[v_{y}(x',y,t) +(v_{x'}(x',y,t))^2]_y\,dx', \ \ t\ge0. 
\eeq
In addition, for any smooth compact support initial condition and any $t>0$, 
the solution develops the constraint 
\begin{equation}
\label{eq:man-cond}
\partial_y {\cal M}(y,t)\equiv 0, \ \ \mbox{where} \ \ 
{\cal M}(y,t)=\int_{-\infty}^{+\infty} \left[v_{y}(x,y,t) +(v_{x}(x,y,t))^2\right]
\,dx.
\end{equation}
identically in $y$ and $t$, but, unlike the  Manakov constraints for the 
Kadomtsev-Petviashvili  (KP) \cite{KP} and for the  dispersionless 
Kadomtsev-Petviashvili  (dKP) \cite{LRT,timman,zobolot} equations, no rapidly 
decaying smooth initial data can satisfy this condition at $t=0$. 

We remark that the problem of non-locality is not typical of integrable 
dispersionless PDEs only, but it is also a generic feature of soliton PDEs 
with 2 spatial variables. Therefore the problem of choosing proper 
integration constants is very important also in the soliton contest, and the 
IST provides the natural choice.  This problem was first posed and discussed 
in \cite{AW} for the KP equation. The final answer for KP was obtained in 
\cite{BPP}, and, later, in \cite{FS}. 

In the remaining part of this introduction we summarize the basic formulas of 
the IST for the Pavlov equation (see, for instance, \cite{GSW}) that will be 
used in this paper.

\subsection{Summary of the IST for the Pavlov equation}

The Pavlov equation is the commutativity condition $[L,M]=0$ for the 
following pair of vector fields: 
\begin{align}
\label{Lax_Pavlov}
L\equiv \partial_y+(\lambda +v_x)\partial_x, \\
M\equiv \partial_t+(\lambda^2+\lambda v_x-v_y)\partial_x.\nonumber
\end{align}
Assuming, as in \cite{GSW}, that the Cauchy datum $v(x,y,0)$ has compact 
support
\begin{equation}
\label{compact}
v(x,y,0)=0 \ \ \mbox{if} \ \ |x|>D_x \ \ \mbox{or} \ \ |y|>D_y,
\end{equation}
is smooth and satisfies some small norm conditions, we define the spectral 
data using the following procedure: 
\begin{enumerate}
\item We define the real Jost eigenfunctions $\varphi_{\pm}(x,y,\lambda)$, 
$\lambda\in\RR$ as the solutions of the eqution
$$
L\varphi_{\pm}(x,y,\lambda)=0,
$$
with the boundary condition:
$$
\varphi_{\pm}(x,y,\lambda)\rightarrow x-\lambda y \ \ \mbox{as} \ \ 
y\rightarrow\pm\infty,
$$
using the correspondent vector fields ODE:
\begin{equation}
\label{ODE}
\frac{dx}{dy}=\lambda+v_x(x,y)
\end{equation}
\item If we denote by $x_{-}(y,\tau,\lambda)$ the solution of (\ref{ODE}) with
the following asymptotics:
$$
x_{-}(y,\tau,\lambda)=\tau+\lambda y + o(1) \ \ \mbox{as} \ \ 
y\rightarrow -\infty,
$$
then the classical time-scattering datum $\sigma(\tau,\lambda)$ is defined 
through the following formula:
\begin{equation}
\label{def_sigma}
\sigma(\tau,\lambda)=\lim\limits_{y\rightarrow+\infty} 
[x_{-}(y,\tau,\lambda)-\tau-\lambda y].
\end{equation}
Equivalently, 
$$
\sigma(\tau,\lambda)=\int_{-\infty}^{\infty} v_x (x_{-}(y,\tau,\lambda),y)dy.
$$
In the linear limit $v\ll 1$ the scattering datum $\sigma(\tau,\lambda)$ 
reduces to the Radon transform of $v_x(x,y)$ \cite{MS4}.
\item 
The spectral data $\chi_{\pm}(\tau,\lambda)$ are defined as the solutions of 
the following shifted Riemann-Hilbert (RH)  problem: 
\beq\label{E:shift-intro}
\sigma(\tau,\lambda)+\chi_{+}(\tau+\sigma(\tau,\lambda),\lambda)-
\chi_{-}(\tau,\lambda)=0,\quad \tau,\lambda \in \RR,
\eeq
where
$\chi_{\pm}(\tau,\lambda)$ are analytic in $\tau$ in the upper and lower 
half-planes $\CC^{\pm}$  respectively, and
$$
\chi_{\pm}(\tau,\lambda)\to 0\ \ \mbox{as}\ \ |\tau|\to\infty.
$$
\end{enumerate}

If the potential $v(x,y,t)$ evolves in $t$ with respect to the Pavlov 
equation, then the scattering and the spectral data evolve in a simple way:
\begin{align}
\label{eq:spectral-evolution}
&\sigma(\tau,\lambda,t)= \sigma(\tau-\lambda^2 t,\lambda,0),\\
&\chi_{\pm}(\tau,\lambda,t)= \chi_{\pm}(\tau-\lambda^2 t,\lambda,0).\nonumber
\end{align}
 
The reconstruction of the potential consists of two steps:
\begin{enumerate}
\item
One solves the following nonlinear integral equation for the time-dependent 
real Jost eigenfunction:
\beq
\label{inversion1}
\psi_-(x,y,t,\lambda)-H_{\lambda}\chi_{-I}\big(\psi_-(x,y,t,\lambda),
\lambda\big)+\chi_{-R}\big(\psi_-(x,y,t,\lambda),\lambda\big)=
x-\lambda y-\lambda^2 t,
\eeq
where $\chi_{-R}$ and $\chi_{-I}$ are the real and imaginary parts of $\chi_-$, 
and $H_{\lambda}$ is the Hilbert transform operator wrt $\lambda$
\beq
H_{\lambda}f(\lambda)=\frac{1}{\pi} \fint\limits_{-\infty}^{\infty}
\frac{f(\lambda')}{\lambda-\lambda'}d\lambda'.
\eeq
In \cite{GSW} it is shown that, for Cauchy data satisfying some explicit 
small-norm conditions, equation (\ref{inversion1}) is uniquely solvable for 
all $t\ge0$. 
\item Once the real time-dependent Jost eigenfucntion is known, the potential 
$v(x,y,t)$ is defined 
by:
\begin{equation}
\label{eq:v-riemann}
v(x,y,t)= -\frac{1}{\pi} \int\limits_{\RR} \chi_{-I}(\psi_{-}(x,y,t,\lambda),
\lambda) d\lambda.
\end{equation}
\end{enumerate}

In addition, in \cite{GSW} it was shown that under the same analytic 
assumptions on the Cauchy data, the function 
$\omega(x,y,t,\lambda)=\psi_-(x,y,t,\lambda)-x+\lambda y + \lambda^2 t$ 
belongs to the spaces $L^{\infty}(d\lambda)$ and $L^{2}(d\lambda)$ for all real 
$x$, $y$ and $t\ge0$, and continuously depends on these variables. Moreover, 
for all $x,y\in\RR$, $t\ge 0$, the following derivatives of 
$\omega$: 
$$
\partial_x \omega, \ \ \partial_y \omega, \ \ \partial_t \omega,  \ \ 
\partial^2_x \omega, \ \ \partial^2_y \omega, \ \ \partial_x\partial_y \omega, 
\ \ \partial_t\partial_x \omega,
$$
are well-defined as elements of the space $L^2(d\lambda)$, they 
continuously depend on $x,y,t$ and are uniformly bounded in 
$\RR\times\RR\times \overline{\RR^+}$.

\section{A lemma from integral geometry and its 
applications}

How is it possible to guess the answer obtained in \cite{GS}? Consider the 
following approximate formula for reconstructing the function $v(x,y,t)$:
\begin{equation}
\label{eq:v-riemann2}
v(x,y,t)\sim -\frac{1}{\pi} \int\limits_{\RR} 
\chi_{-I}(\psi^{(0)}(x,y,t,\lambda),\lambda) d\lambda, \ \ \mbox{where} \ \ 
\psi^{(0)}(x,y,t,\lambda)=x-\lambda y -\lambda^2 t.
\end{equation}
In this approximation we replace the term $\omega(x,y,t,\lambda)$ by $0$ in 
formula (\ref{eq:v-riemann}); this approximation is valid for $|x|\gg1$, 
because in this region the wave function can be approximated by its 
normalization.

Let us apply the following result from integral geometry (the authors were 
not able to find this result in the literature).

\begin{figure}[h]
\begin{center}
\includegraphics[height=4cm]{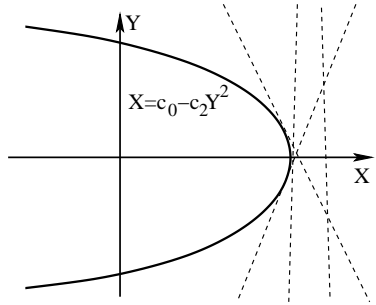}
\end{center}
\caption{\label{fig:1} Integration contour for $I_p(c_0,c_2)$ (parabolic) and 
integration lines for $I_l(a,b)$ (dashed).}
\end{figure}

\begin{lemma}
\label{lem:1}
Assume that we have a sufficiently good function $f(X,Y)$ in the real plane 
and a parabola $X=c_0 -c_2 Y^2$. Consider the following partial Radon data:
\begin{equation}\label{partial Radon data 1}
I_l(a,b)=\int\limits_{-\infty}^{+\infty} f(a+b Y,Y)dY,
\end{equation}
where only lines $X=d_0+d_1 Y$, not intersecting the parabola 
$X=c_0 -c_2 Y^2$, are taken (but the lines are permitted to be tangent to 
parabola), see Figure~\ref{fig:1}. Then the integral 
\begin{equation}
I_p(c_0,c_2)=\int\limits_{-\infty}^{+\infty} f(c_0-c_2 Y^2,Y)dY 
\end{equation}
can be expressed in terms of the data (\ref{partial Radon data 1}) by 
inverting the Melling transform (see formula (\ref{eq:tom10})).
\end{lemma}

\begin{figure}[h]
\begin{center}
\includegraphics[height=4cm]{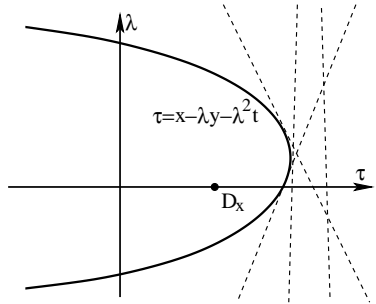}
\end{center}
\caption{\label{fig:2} Integration contours for $t>0$ (parabolic) and for 
$t=0$ (dashed).}
\end{figure}

Assume now that $t>0$ and $x>D_x$. Using the fact that $v(x,y,0)=0$ for 
$x>D_x$ we see that integrating $\chi_{-I}(\tau,\lambda)$ over any straight 
line not intersecting the parabola $x-\lambda y - \lambda^2 t$ is equal to 
zero (up to the above approximation error), see Figure~\ref{fig:2}. 
To construct $v(x,y,t)$ in our approximation, we have to calculate  
the integral of $\chi_{-I}(\tau,\lambda)$ on the parabola, which is 0 by 
Lemma~\ref{lem:1}, implying that $v_t$, $v_x$, $v_y$ are also zero for large 
positive $x$ and $t$ positive. Therefore, from equation  (\ref{Pavlov-ev}), 
it follows that $\partial_x^{-1}=-\int_x^{+\infty}$.

Now we would like to consider a completely different application of 
Lemma~\ref{lem:1}. Indeed a similar lemma, in which the parabola is replaced 
by the circle, is already known in the literature \cite{Cormack1}, and finds 
applications in tomography.  

Assume that we have a two-dimensional linear tomographic problem with a 
convex opaque obstacle. In this situation, only the integrals along the 
straight lines non-intersecting the obstacle are known (the partial Radon 
data). If $f(X,Y)$ belongs to the Schwartz class, its reconstruction from the 
partial Radon data is unique, due to the hole theorem, but severely ill-posed 
\cite{Natterer}. In the class of potentials  $f(X,Y)$ with polynomial decay, 
we do not have unique solvability of the inverse problem (the reconstruction 
of $f(X,Y)$) anymore. Consider, f.i., the function 
$$
f(X,Y)=\Re\left(\frac{1}{(X-X_0+i(Y-Y_0))^n} \right), \ \ n\ge 2,
$$
where the point $(X_0,Y_0)$ is located inside the obstacle. For all straight 
lines not intersecting the obstacle, the integral of $f(X,Y)$ along these 
lines is equal to 0 because $f(X,Y)$ is harmonic; moreover the function 
$f(X,Y)$ is well-localized for large $n$. It means that, if $f(X,Y)$ is a 
more general potential, its harmonic part does not contribute to the partial 
Radon data, i.e. it belongs to the kernel of the direct Radon transform 
\cite{Natterer}. Therefore, for the problem with obstacle, it is natural not 
to try to reconstruct the complete potential, but only some integral 
characteristics, like the integral over the boundary of the obstacle. The 
measure of the integral over the boundary of the obstacle should have 
the following property: for each function harmonic outside the obstacle with 
sufficient localization the integral should be equal to zero. 

Let us discuss the cases of elliptic and parabolic obstacles, for wich it is 
indeed possible to express the integral over the boundary of the obstacle in 
terms of the partial Radon data.  
\begin{itemize}
\item The obstacle is bounded by the oval 
$\frac{x^2}{\alpha^2}+\frac{y^2}{\beta^2}=1$, i.e. the forbidden area is 
defined by the inequality  $\frac{x^2}{\alpha^2}+\frac{y^2}{\beta^2}\le 1$.
\item The obstacle is bounded by the parabola $x=-c y^2$, i.e. the forbidden 
area is: $x+cy^2\le0$. 
\end{itemize}

In the first case we would like to calculate the following integrals:
\begin{equation}
\label{eq:tom1}
I_e(r) =\frac{1}{r}\int\limits_{x^2/\alpha^2+y^2/\beta^2=r^2} f(x,y)
\sqrt{\frac{dx^2}{\alpha^2}+\frac{dy^2}{\beta^2}}, \ \ 
r\ge 1.
\end{equation}
through the partial Radon data. Since, through the rescaling 
$\tilde x=x/\alpha$,  $\tilde y=y/\beta$, the integral (\ref{eq:tom1}) can be 
reduced to the case of a circular obstacle:
\begin{equation}
\label{eq:tom2}
I_e(r) =\frac{1}{r}\int\limits_{\tilde x^2+\tilde y^2 =r^2} f(\tilde x,\tilde y)ds, 
\ \ ds^2=d\tilde x^2+d\tilde y^2, \ \ r\ge1 ,
\end{equation}
we can use the following lemma form integral geometry \cite{Cormack1}:

\begin{lemma}
\label{lem:2}
Assume that we have a sufficiently good function $f(x,y)$ in the real plane 
and the circle $x^2+y^2=r^2$. Then the integral 
\begin{equation}
I_c(r)=\frac{1}{r} \int\limits_{x^2+y^2=r^2} f(x,y)ds, \ \  ds^2=d x^2+d y^2,
\end{equation}
can be expressed in terms of the partial radon data:
\begin{equation}
I_l(r_1,\phi)=\int\limits_{l(r_1,\psi)} f(x,y)ds, \ \ r_1\ge r,
\end{equation}
where $l(r_1,\phi)$ denotes the straight line such that $r_1$ is the distance 
between it and the origin, and $\phi$ denotes the angle between the line and 
the origin. Again, only the lines not intersecting the obstacle are considered.
\end{lemma}

In the second case the natural integral characteristics are:
\begin{equation}
\label{eq:tom4}
I_p(c_0) =\int\limits_{x+cy^2=c_0} f(x,y)dy, \ \ c_0>0,
\end{equation}
and we use Lemma~\ref{lem:1}. 

We expect that Lemmas~\ref{lem:1} and \ref{lem:2} can be generalized to a 
larger class of convex obstacles.

\section{Proofs of the Lemmas}

{\bf Proof of Lemma~\ref{lem:1}:} Without loss of generality we may assume 
$c_2=1$. The line $x=a+b y$ does not intersect the parabola iff 
$(a,b)\in U(c_0)$, where
$$
U(c_0)=\left\{(a,b)\, \bigg |\, a\ge \frac{b^2}{4}+c_0\right\},
$$
Let us calculate $J(c_0)=\iint\limits_{U(c_0)} I_l(a,b)\, d a\, d b$.

$$
J(c_0)=\iint\limits_{U(c_0)} I_l(a,b)\, d a\, d b= \iint\limits_{U(c_0)} \, d a\, 
d b \int\limits_{-\infty}^{+\infty}dy f(a+by,y)=
$$
$$
= \int\limits_{-\infty}^{+\infty}dy \iint\limits_{a-c_0\ge b^2/4}da\, d b\, f(a+by,y)
$$
Denote $\tau=a+b y$, $da db = d\tau db$. then
$$
J(c_0)= \int\limits_{-\infty}^{+\infty}dy  \iint\limits_{\tau-c_0\ge b^2/4+b y}d\tau\, 
d b\, f(\tau,y)=
$$
$$
=\int\limits_{-\infty}^{+\infty}dy  \int\limits_{\tau\ge c_0-y^2}d\tau 
\int\limits_{b^2/4+b y\le \tau-c_0 } d b\, f(\tau,y)=
\int\limits_{-\infty}^{+\infty}dy  \int\limits_{\tau\ge c_0-y^2}d\tau f(\tau,y) 
\int\limits_{b^2/4+b y\le \tau-c_0 } 1\, d b=
$$
$$
=
\int\limits_{-\infty}^{+\infty}dy  \int\limits_{\tau\ge c_0-y^2}d\tau 
\left[f(\tau,y)\, 4\sqrt{y^2+\tau-c_0}\right].
$$
Denote $\tau=\tau'-y^2$; then
\begin{align}
\label{eq:tom10}
&J(c_0)= 4\int\limits_{-\infty}^{+\infty}dy  \int\limits_{\tau'\ge c_0}d\tau' 
\left[f(\tau'-y^2,y)\, \sqrt{\tau'-c_0}\right]=
4 \int\limits_{\tau'\ge c_0}d\tau' \sqrt{\tau'-c_0}  
\int\limits_{-\infty}^{+\infty}dy f(\tau'-y^2,y)=\\
& = 4  \int\limits_{\tau'\ge c_0}d\tau' \sqrt{\tau'-c_0}\, I_p(\tau');\nonumber
\end{align}
Therefore our problem is reduced to the inversion of the Abel transform 
\cite{Abel}:
\beq
I_p(\tau)=\frac{1}{2\pi}\frac{d}{d\tau}\int\limits_{\tau}^{\infty}\frac{J'(c_0)}{\sqrt{c_0-\tau}}dc_0 .
\eeq

{\bf Proof of Lemma~\ref{lem:2}:} 

\begin{center}
\includegraphics[width=4cm]{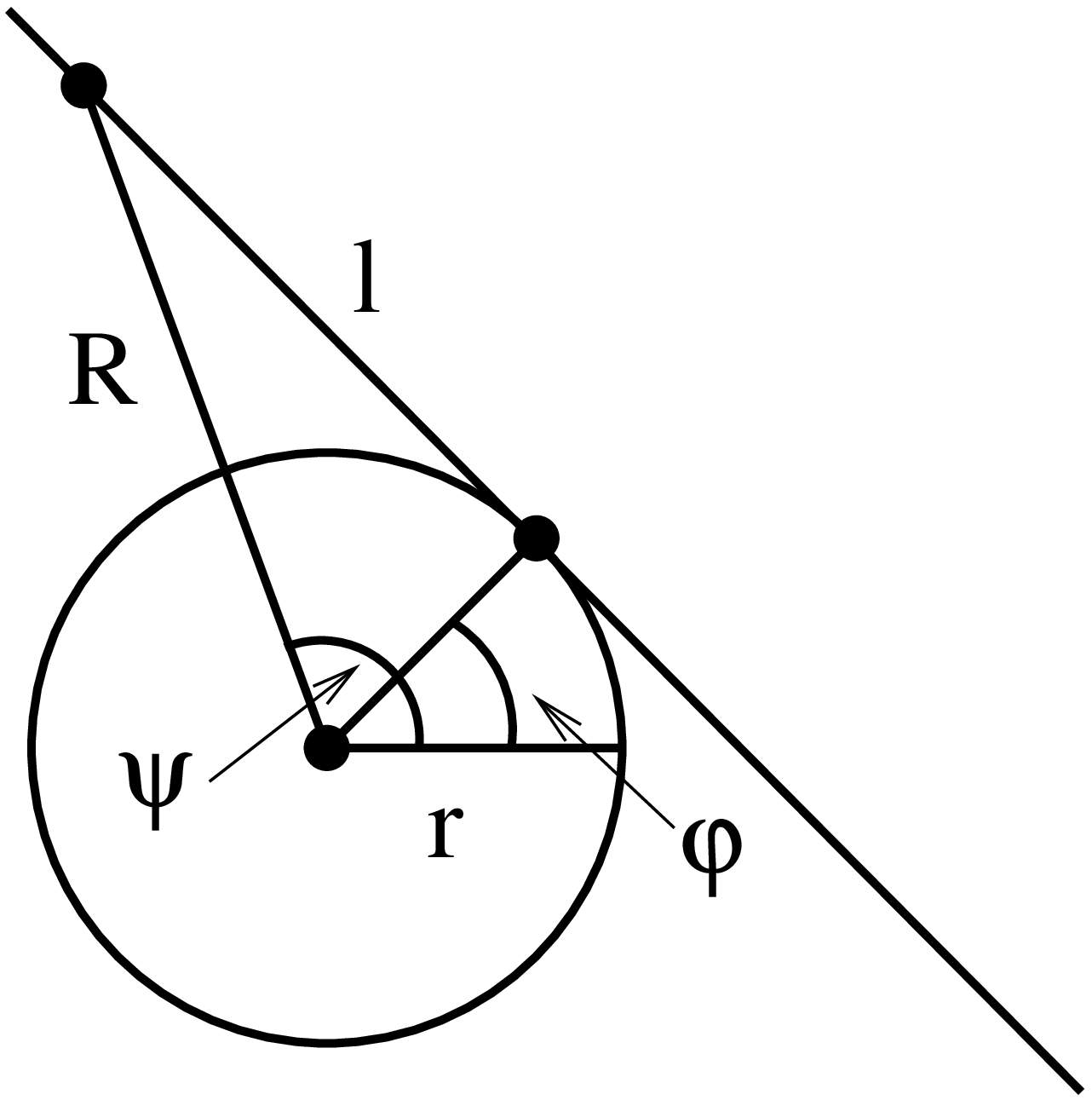}
\end{center}
Let us parametrize the lines in the plane by two parameters $(r,\phi)$, where 
$r$ denotes the distance from the origin and $\phi$ denotes the angle 
between the $x$-axis and the perpendicular form the origin to the line. 

Denote by $l$ the coordinate on the line, $l=0$ at the perpendicular. Denote 
by $(R,\psi)$ the polar coordinates of a point of the line. We have the 
following change of coordinates:
\begin{equation}
\left\{\begin{array}{l} R=\sqrt{r^2+l^2} \\ \psi=\phi+\arctan\left(\frac{l}{r}
\right)  
\end{array}\right.
\end{equation}
Assume that $r$ is fixed. 
$$
d\psi\wedge dR=
\det\left|\begin{array}{cc} \frac{\partial\psi}{\partial\phi} & 
\frac{\partial\psi}{\partial l}  \\
\frac{\partial R}{\partial\phi} & \frac{\partial R}{\partial l} \end{array} 
\right|d\phi\wedge dl=
\det\left|\begin{array}{cc} 1 & \frac{r}{r^2+l^2}  \\
0 & \frac{l}{\sqrt{l^2+r^2}} \end{array} \right|d\phi\wedge dl=
\frac{l}{\sqrt{l^2+r^2}} d\phi\wedge dl,
$$
and
$$
d\phi\wedge dl = \frac{R}{\sqrt{R^2-r^2}}d\psi\wedge dR
$$

Let us denote by $I_l(r,\phi)$ the integral
$$
I_l(r,\phi)= \int\limits_{-\infty}^{\infty}dl f(x(l,\phi),y(l,\phi)) dl
$$
Consider the following integral: 
$$
J(r)=\int\limits_{0}^{2\pi}d\phi I_l(r,\phi) = \int\limits_{0}^{2\pi}d\phi 
\int\limits_{-\infty}^{\infty}dl f(x(l,\phi),y(l,\phi)) =
$$
$$
=\int\limits_{0}^{2\pi}d\psi \int\limits_{r}^{\infty}dR f(R\cos\psi,R\sin\psi) 
\frac{R}{\sqrt{R^2-r^2}} =
\int\limits_{r}^{\infty} I_c(R) \frac{R}{\sqrt{R^2-r^2}} dR,
$$
where 
$$
I_c(R)=\int\limits_{0}^{2\pi}  f(R\cos\psi,R\sin\psi) d\psi.
$$
Therefore our problem is reduced again to the inversion of the Abel transform:
\beq
I_c (R)=-\frac{2}{\pi}\int\limits_R^{\infty}\frac{J'(r)}{\sqrt{r^2-R^2}}dr .
\eeq

Acknowledgments: The authors would like to express their gratitude to 
R.G.~Novikov for consultations about tomographic results. 
The first author was partially supported by the Russian 
Foundation for Basic Research, grant 13-01-12469 ofi-m2, by the program 
``Leading scientific schools'' (grant NSh-4833.2014.1), by the program 
``Fundamental problems of nonlinear dynamics'', Presidium of RAS, by the INFN 
sezione di Roma, and by the PRIN 2010/11 No~JJ4KPA\_004 of Roma 3.

\end{document}